\documentclass[10pt,onecolumn,letterpaper]{article}

\usepackage[pagenumbers]{cvpr} 

\usepackage{braket}
\usepackage{multirow}
\usepackage{algorithmic}
\usepackage{algorithm}
\usepackage{bbding}

\definecolor{cvprblue}{rgb}{0.21,0.49,0.74}
\usepackage[pagebackref,breaklinks,colorlinks,allcolors=cvprblue]{hyperref}

\def\paperID{30119} 
\def\confName{CVPR}
\def\confYear{2026}

\title{Fidelity-Preserving Quantum Encoding for Quantum Neural Networks}

\author{
Yuhu Lu\\
School of Computer Science and Engineering, Central South University\\
Changsha 410083, China\\
\and
Jinjing Shi $^{*}$\\
School of Electronic Information, Central South University\\
Changsha 410083, China\\
{\tt\small shijinjing@csu.edu.cn}
}

\begin{document}
\maketitle

\begin{abstract}
Efficiently encoding classical visual data into quantum states is essential for realizing practical quantum neural networks (QNNs).
However, existing encoding schemes often discard spatial and semantic information when adapting high-dimensional images to the limited qubits of Noisy Intermediate-Scale Quantum (NISQ) devices.
We propose a Fidelity-Preserving Quantum Encoding (FPQE) framework that performs near lossless data compression and quantum encoding.
FPQE employs a convolutional encoder-decoder to learn compact multi-channel representations capable of reconstructing the original data with high fidelity, which are then mapped into quantum states through amplitude encoding.
Experimental results show that FPQE performs comparably to conventional methods on simple datasets such as MNIST, while achieving clear improvements on more complex ones, outperforming PCA and pruning based encodings by up to 10.2\% accuracy on Cifar-10.
The performance gain grows with data complexity, demonstrating FPQE's ability to preserve high-level structural information across diverse visual domains.
By maintaining fidelity during classical to quantum transformation, FPQE establishes a scalable and hardware efficient foundation for high-quality quantum representation learning.
\end{abstract}

\section{Introduction}
\label{sec:intro}

Quantum computing has emerged as a promising paradigm for advancing machine learning and visual understanding, offering the potential to exploit the expressive power of quantum states for high-dimensional data analysis \cite{Shor,Grover}. With recent progress in quantum hardware and simulation frameworks, quantum neural networks (QNNs) have been increasingly explored for classification, generative modeling, and representation learning
Despite rapid progress, one key bottleneck remains unresolved is efficiently encoding classical visual data into quantum states without losing critical spatial or semantic information \cite{QAOA2, QAOA3,xiong2024circuit}.
This problem becomes even more severe under the constraints of Noisy Intermediate-Scale Quantum (NISQ) devices, where the number of available qubits is limited and noise sensitivity is high.

Existing quantum encoding strategies primarily fall into three categories.
The first type directly maps pixel or feature vectors to quantum states via amplitude \cite{gonzalez2024efficient,pagni2025fast} or angle encoding \cite{ovalle2023quantum}.
While these methods are straightforward, but they may struggle to handle high-dimensional data, leading to partial loss of local structures and correlations.
The second type applies classical dimensionality reduction like principal component analysis (PCA) \cite{lloyd2014quantum} and autoencoder compression \cite{huang2020realization} before quantum encoding.
Although this reduces qubit requirements, it introduces irreversible information loss, weakening the representational fidelity and degrading performance on complex vision datasets.
More recent quantum oriented encoders including patch-based Single Qubit Encoding (SQE) \cite{easom2022efficient} and pruning-based adaptive threshold pruning (ATP) \cite{afane2025atp} address certain resource constraints but still degrade rapidly on datasets like Cifar-10. These methods tend to preserve pixel level statistics while failing to retain global or local structure, which we find to be essential for quantum models to learn meaningful decision boundaries.

The gap between compressibility and fidelity preservation motivates the need for encoding methods that remain compatible with tight qubit budgets while retaining the structure necessary for visual recognition \cite{kaufmann2018high,liu2024fidelity}. This requirement is rarely stated explicitly, yet our experiments show that structural fidelity that captured by metrics such as SSIM is strongly correlates with QNN performance, whereas commonly used metrics such as MSE and PSNR do not reliably predict learning outcomes \cite{balasubramani2025quantum}. This observation suggests that the primary bottleneck in existing encoding methods is not dimensionality reduction itself but the loss of structural information during the transformation process.

To overcome these limitations, we propose Fidelity-Preserving Quantum Encoding (FPQE), a preprocessing framework that achieves near-lossless classical-to-quantum data transformation.
FPQE employs a convolutional encoder-decoder that learns low-dimensional, multi-channel representations capable of reconstructing the original image with high fidelity.
The compressed representations are then mapped into quantum states using amplitude encoding, ensuring that both global and local structures are preserved during the transformation process.
Unlike conventional methods that optimize only for compression or feasibility, FPQE explicitly optimizes for fidelity preservation, thereby improving the quality of quantum representations available to downstream QNNs.

We evaluate FPQE on three benchmark datasets: MNIST, FashionMNIST, and Cifar-10.
While FPQE performs comparably to baseline encoders on simple datasets such as MNIST, it achieves up to 10.2\% higher accuracy on more complex datasets like Cifar-10, indicating that its advantage grows with data complexity.
This improvement demonstrates FPQE's ability to retain rich structural correlations that conventional dimensionality-reduction techniques fail to preserve.

The main contributions of this work are summarized as follows:
\begin{enumerate}
  \item We propose a novel Fidelity-Preserving Quantum Encoding (FPQE) that minimizes information loss during classical-to-quantum transformation.
  \item We build quantum neural networks powered by FPQE and conduct extensive binary classification experiments on MNIST, FashionMNIST, and Cifar-10. Across all datasets, FPQE consistently outperforms traditional encoding schemes, with the performance gap widening as visual complexity increases.
  \item We provide a comprehensive fidelity analysis using MSE, PSNR, and SSIM, showing that FPQE maintains not only pixel-level similarity but also high structural fidelity. This structure preservation directly translates into improved downstream QNN performance.
\end{enumerate}
\section{Renalted Works}
\label{sec:related}
\textbf{Classical to Quantum Encoding: }
Research on quantum machine learning has progressed rapidly in recent years, driven by advances in quantum hardware and the development of variational quantum algorithms. Within this landscape, one of the most critical challenges is the design of efficient encoding mechanisms that translate classical data into expressive quantum states suitable for downstream learning \cite{weigold2020data,rath2024quantum,schuld2021effect}. 

The most widely used strategies include amplitude encoding, where classical feature vectors are normalized and embedded into the amplitudes of a quantum state; angle encoding, where data values modulate qubit rotation angles; and basis encoding, where each bit of classical data is mapped to the state of a qubit \cite{schuld2020circuit,benedetti2019parameterized}.
While conceptually simple, these approaches scale poorly for high-dimensional visual inputs, as the number of required qubits increases exponentially with image size.
Moreover, when operating under the constraints of Noisy Intermediate-Scale Quantum (NISQ) devices \cite{preskill2018quantum}, direct high-dimensional encoding introduces noise and decoherence that severely degrade the representational fidelity of the resulting quantum states.

To mitigate the dimensionality mismatch between high-resolution images and the small number of qubits available on NISQ devices, several works explore classical dimensionality reduction techniques before encoding the data into quantum states. Principal component analysis and linear projections \cite{lloyd2014quantum} reduce dimensionality by capturing global variance, but they fail to preserve fine grained spatial patterns. Autoencoder based strategies offer more flexible representations, yet they often optimize reconstruction error without explicitly maintaining structural or semantic fidelity. As a result, these methods may yield compact encodings but do not necessarily produce quantum relevant representations that benefit downstream variational models.

More recent studies propose learning based or algorithmic compression schemes specifically designed for quantum applications. SQE transforms small image patches into single qubit states and has been shown to improve the efficiency of quantum image classification \cite{easom2022efficient}. However, SQE focuses on local patch embeddings and only partially preserves global semantic structures, making it less effective on visually complex datasets with high inter class variability. ATP introduces a pruning mechanism to reduce redundant coefficients prior to quantum encoding, enabling more efficient state preparation, but it similarly sacrifices structural information when the underlying images exhibit rich textures and global patterns \cite{afane2025atp}. Both methods illustrate the importance of balancing compression and representational fidelity, yet they provide limited strategies for preserving overall visual structure.

To alleviate these issues, various preprocessing based encoding schemes have been proposed.
PCA and linear embeddings are among the earliest solutions, projecting data into a lower-dimensional subspace before quantum encoding \cite{perez2021training,broughton2020tensorflow,poduval2024hdqmf}.
Although effective for dimensionality reduction, PCA often destroys spatial correlations essential for visual recognition.
Alternatively, several works explore autoencoder based quantum data compression, where classical or hybrid neural networks are trained to generate compact latent vectors suitable for amplitude encoding \cite{chen2021hybrid,liao2022quantum}.
Another strategy is block or patch wise encoding, in which images are partitioned into smaller subregions, each independently encoded as a quantum state \cite{havlicek2019supervised,benedetti2019parameterized}.
These methods reduce qubit requirements but lose inter patch dependencies, limiting their ability to represent global structure.\\
\textbf{Quantum Neural Networks: }
QNNs aim to leverage the expressive capacity of parameterized quantum circuits (PQCs) for learning tasks. Unlike classical networks, which operate in Euclidean space, QNNs manipulate data within exponentially large Hilbert spaces, enabling richer nonlinear transformations and potentially more compact models \cite{schuld2015introduction,abbas2021power,yang2022semiconductor,doan2022hybrid}. Early works demonstrated that PQCs could serve as universal function approximators, capable of modeling highly complex decision boundaries with relatively few trainable parameters \cite{schuld2019quantum,sim2019expressibility}. Subsequent studies explored their application to supervised learning \cite{mitarai2018quantum}, generative modeling, and representation learning, confirming the flexibility of variational quantum circuits for extracting patterns not easily captured by classical models.

A growing number of hybrid architectures combine classical feature extraction with quantum classifiers. For example, \cite{perez2021training} integrates convolutional neural networks with PQCs to handle high-dimensional images, while \cite{broughton2020tensorflow} introduces a unified classical-quantum training framework enabling end-to-end optimization. Fully quantum architectures such as quantum convolutional neural networks (QCNNs) mimic the hierarchical structure of classical CNNs but process information entirely in quantum space. These developments demonstrate the promise of quantum models for visual tasks and suggest that improved quantum representations could further enhance performance.

Recent studies show that QNNs can approximate complex nonlinear mappings using fewer parameters than classical models, thanks to high-dimensional Hilbert space representations \cite{sim2019expressibility,mitarai2018quantum}.
However, these benefits depend heavily on the quality of data encoding, especially the low fidelity mappings cause state overlap, reducing the separability of quantum features and diminishing learning performance.

Our work addresses this overlooked gap by placing fidelity preservation at the center of the classical to quantum transformation process. The proposed FPQE framework introduces a learnable convolutional encoder-decoder trained explicitly to retain both pixel level appearance and structural composition before quantum state preparation. Unlike traditional dimensionality reduction techniques that prioritize qubit feasibility, FPQE is optimized to maintain global shapes, local textures, and semantic boundaries, which is essential for robust quantum learning.

Once trained, FPQE produces multi-channel low-dimensional representations that align with the dimensional constraints of amplitude encoding while preserving the structure necessary for effective discrimination. By maintaining high structural similarity, FPQE reduces state overlap in Hilbert space and improves the separability of quantum features, enabling QNNs to better exploit their expressive capacity. Experimental results across MNIST, FashionMNIST, and Cifar-10 show that FPQE yields consistently stronger performance than angle encoding, amplitude encoding, PCA, SQE, and ATP, with the performance advantage growing as data complexity increases.
\section{Methods}
\label{sec:method}
\begin{figure*}
  \centering
  \includegraphics[width=0.9\textwidth]{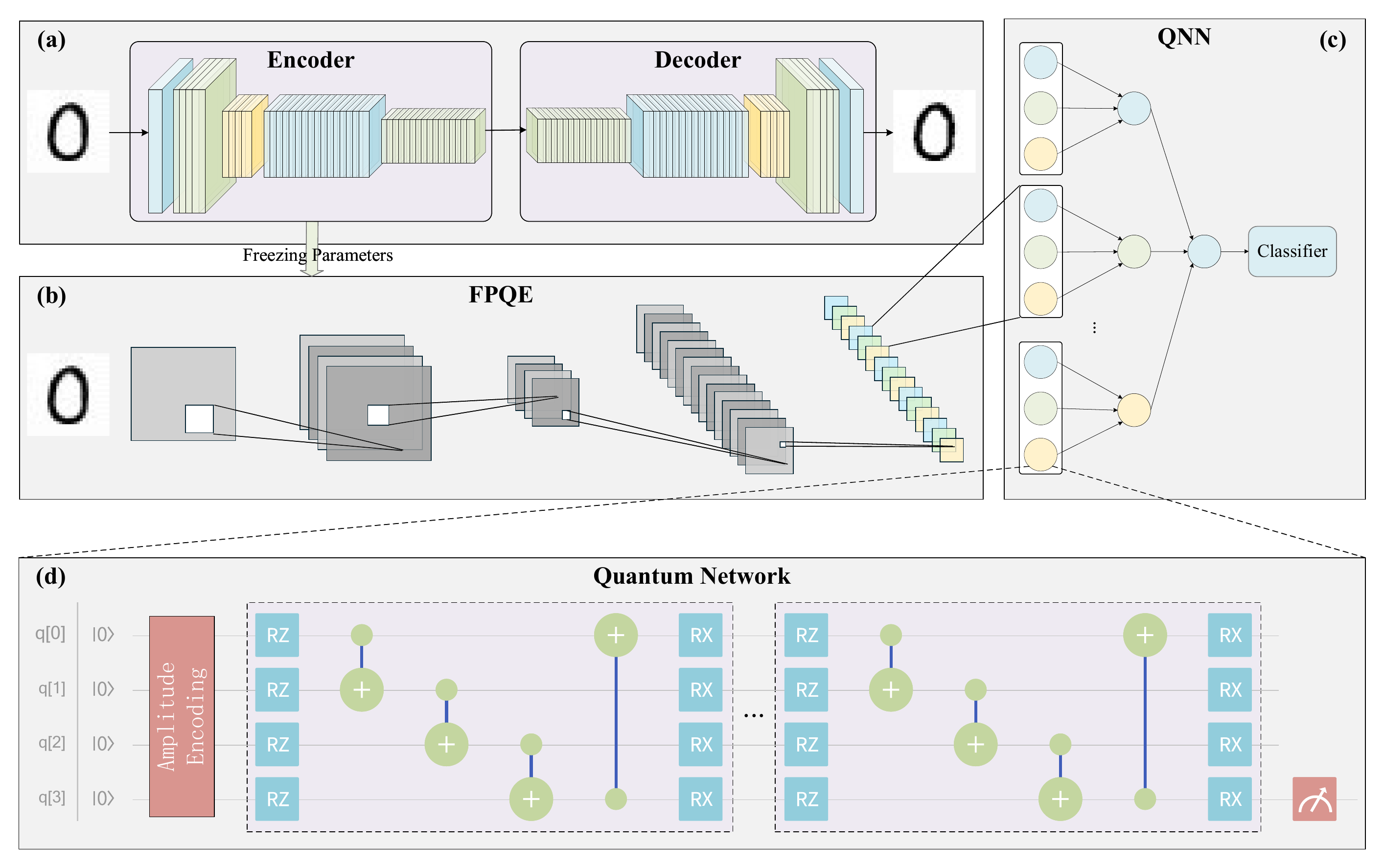}
  \caption{FPQE framework}
  \label{FPQE_framework}
\end{figure*}

The proposed FPQE is designed to efficiently map high-dimensional classical data into quantum states with minimal information loss. The framework integrates classical convolutional feature extraction with quantum data encoding, allowing quantum neural networks to operate effectively on NISQ devices. The proposed FPQE based QNNs framework is shown if Fig. \ref{FPQE_framework}, which consists of two main stages: 
\begin{enumerate}
  \item A classical fidelity-preserving encoder trained through an encoder–decoder reconstruction objective is shown in Fig. \ref{FPQE_framework}(a, b);
  \item A downstream QNN that consumes the compressed multi-channel representation is shown in \ref{FPQE_framework}(c,d).
\end{enumerate}
\subsection{Encoder-Decoder for Fidelity Preservation}
The encoder-decoder structure is shown in \ref{FPQE_framework}(a).
The encoder architecture follows a convolutional compression paradigm with repeated Conv-BN-ReLU-Pooling layers that progressively reduce spatial dimensions while increasing channel depth.
This design allows encoder to aggregate local texture and edge information into multi-channel latent codes that maintain the image's global and local structure.
The decoder mirrors this structure using transposed convolutions and upsampling layers, ensuring near lossless reconstruction.

Formally, given an input image $x \in \mathbb{R}^{C\times H\times W}$, the encoder $E_\theta(\cdot )$ projects it into a compact latent representation 
\begin{equation}
  z = E_\theta(x)\in\mathbb{R}^{c\times h\times w},
\end{equation}
where $h<H$ and $w<W$. 
The decoder $D_\eta(\cdot)$ mirrors the encoder using transposed convolutions and upsampling layers to reconstruct the input 
\begin{equation}
  x'=D_\eta(z). 
\end{equation}
The encoder-decoder is trained to minimize reconstruction loss:
\begin{equation}
  \mathcal{L}_{rec} = \text{MSE}(x,x^{'})
\end{equation}
to ensure that $z$ retains the essential spatial and semantic information of $x$.
Through this training strategy, the encoder learns a compressed feature representation that is compact enough to match quantum hardware constraints while retaining the structural semantics needed for classification.

\subsection{FPQE structure}
Once the encoder-decoder has converged, we discard the decoder and freeze the encoder weights, which is shown in \ref{FPQE_framework}(b), forming the FPQE:
\begin{equation}
  z = E_{FPQE}(x)\in \mathbb{R}^{c\times h\times w}
\end{equation}
FPQE maps the original high-dimensional image to a multi-channel and low-dimensional tensor that is directly compatible with amplitude encoding. Because the encoder was trained with structural fidelity constraints, the resulting representation maintains global shape, edges, and textural information that traditional classical reduction methods fail to preserve.

The output tensor is flattened and normalized:
\begin{equation}
  v = \text{flatten(z)}, \psi = \frac{v}{\| v \Vert _2}, \psi \in \mathbb{R}^{c\times (h\times w)}
\end{equation}
where $\psi$ becomes the amplitude vector of the quantum state. This stage ensures that quantum circuits operate on structurally meaningful, fidelity-preserved data instead of heavily distorted encodings.

\subsection{FPQE based Quantum neural network}
We construct a quantum neural network that utilizes only quantum parameters. The overall framework is illustrated in Fig. \ref{FPQE_framework}(c,d), where the input is processed using FPQE. Note that the circle in \ref{FPQE_framework}(c) represents a parametrized quantum circuit, as shown in Fig. \ref{FPQE_framework}(d). For classification tasks, the quantum network is generally structured with multiple layers; the measurement outcomes from a preceding layer serve as inputs to the subsequent layer, and the final measured output is used as the basis for classification decisions and for constructing the loss function. The designed quantum neural network utilizes a small number of qubits while encompassing all trainable parameters of the entire network, thereby realizing a neural network model that is fully parameterized by quantum parameters.
The process of obtaining the classification result from the previously obtained $\psi$ is as follows.\\
\textbf{Quantum state mapping: }
The Quantum state mapping is the amplitude encoding in \ref{FPQE_framework}(d).
For each channel $\psi_k \in \mathbb{R}^{(h\times w)}$ of the latent tensor is normalized and encoded into quantum state $\ket{\psi_k}$ using amplitude encoding:
\begin{equation}
  \ket{\psi_k} = \sum_{i=1}^{h*w}\psi_{k,i}\ket{i}.
\end{equation}
Then the encoded input can be represented as $\ket{\psi}=\{\ket{\psi_1}, \ket{\psi_2}, \ldots , \ket{\psi_c}\}$.
Note that the encoded $\ket{\psi}$ uses only $\log (w\times h)$ qubits, which is logarithmic degree of $\psi_k$.\
\textbf{Parameterized quantum circuit: }
We define the parameterized quantum circuit in Fig. \ref{FPQE_framework}(d) as $U_{\Theta, k}$, which can process the encoded quantum state $\ket{\psi_k}$ to:
\begin{equation}
  \ket{\phi_k} = U_{\Theta, k}\ket{\psi_k}.
\end{equation}
The parameterized result $\varphi_k = \braket{\psi_k\|Z\|\psi_k}$ is measured using Pauli Z measurement.
Then the result grained after the first layer is:
\begin{equation}
  \varphi=\{\varphi_1, \varphi_2,\ldots,\varphi_c\} \in \mathbb{R}^{c}.\
\end{equation}
\textbf{Repeat and training: }
By repeating the quantum network (including quantum state mapping and parameterized quantum circuit) for fixed layer $L$, the output of QNN $y'$ is obtained for classification using CrossEntropy.
The overall process of FPQE based QNN is subscribed in Algorithm \ref{qnn_alg}.
\begin{algorithm}
\caption{FPQE based QNN}
\label{qnn_alg}
  \begin{algorithmic}[1]
  \REQUIRE Input data $x$, label $y$, encoder $E_\theta$, decoder $D\eta$, parameterized quantum circuit $U_\Theta$, QNN layer $L$.
  \ENSURE $x\in\mathbb{R}^{C\times H\times W}$

    \vspace{3pt}
    \STATE \textbf{Step 1: Encoder-Decoder}
    \STATE $z = E_\theta(x)$
    \STATE $x' = D_\eta(z)$
    \STATE $\mathcal{L}_{rec} = \text{MSE}(x,x^{'})$
    \STATE train $E_\theta$ and $D_\eta$ using $\mathcal{L}_{rec}$

    \vspace{3pt}
    \STATE \textbf{Step 2: FPQE}
    \STATE Freezing the parameters $\theta$ of $E_\theta$
    \STATE $E_{FPQE} \gets E_\theta$
    \STATE $z = E_{FPQE}(x)$
    \STATE $v = \text{flatten(z)}, \psi = \frac{v}{\| v \Vert _2}$

    \vspace{3pt}
    \STATE \textbf{Step 3: QNN}
    \FOR{$l \text{ in } L$}
      \STATE $\ket{\psi} = \{\ket{\psi_k} = \sum\psi_{k,i}\ket{i} \text{for k in range}(\text{len}(\psi))\}$
      \STATE $\ket{\phi} = \{U_{\Theta, k}\ket{\psi_k}   \text{for k in range} (\text{len}(\ket{\psi}))\}$
      \STATE $\varphi = \braket{\psi\|Z\|\psi}$
    \ENDFOR
    \STATE $y' \gets \varphi$
    \STATE $\mathcal{L}_{QNN} = \text{CrossEntropy}(y', y)$
    \STATE train $U_{\Theta}$ using $\mathcal{L}_{QNN}$
    \vspace{3pt}
  \STATE \textbf{Output:} Return argmax($y'$, dim=1) as the classification results.

  \end{algorithmic}
\end{algorithm}




\section{Experiments}
\subsection{Steup}
We evaluate FPQE on three standard visual benchmarks: MNIST, FashionMNIST, and Cifar-10, each reduced to a binary classification task.
The encoder-decoder is implemented with three convolutional blocks (kernel size 3, stride 2) followed by ReLU activations and batch normalization, producing a latent representation of size , where 
(h,w,c) is adjusted per experiment.
The decoder mirrors this structure using transposed convolutions.
Training is performed using the Adam optimizer with a learning rate of 
 for 100 epochs.
After training, the encoder is frozen, and its output channels are mapped into quantum states via amplitude encoding.

\subsection{Encoding and Preprocessing Methods}
We conducted a comparative experiment of FPQE against various encoding methods, including angle encoding, amplitude encoding, PCA, SQE, and AQT. The configurations for each encoding method are shown in Tab. \ref{resources}.
\begin{table}[htbp]
  \begin{center}
    \caption{The configurations for different encoding method.}
    \label{resources}
    \begin{tabular}{cccc}
      \hline
      & Qubits & z.shape & pruning \\
      \hline
      Angle & 9 & (3,3) & - \\
      Amplitude & 8 & (16,16) & - \\
      PCA & 9 & (9) & - \\
      SQE & 9 & (3,9) & - \\
      AQT & 9 & (16,16) & \Checkmark \\
      FPQE & 6 & (64,64) & -\\
      \hline
    \end{tabular}
  \end{center} 
\end{table}

\subsection{Comparison Results}
Tab.\ref{exp_results} presents the binary classification accuracy of different encoding methods across MNIST, FashionMNIST, and Cifar-10 datasets.
For each dataset, we compared the performance of FPQE against several baseline methods, including Angle Encoding, Amplitude Encoding, PCA, SQE, and ATP.
\begin{table*}[htbp]
  \begin{center}
    \caption{The binary classification accuracy (\%) of different encoding methods.}
    \label{exp_results}
    \begin{tabular}{c|ccccc|cccc|c}
      \hline
      & \multicolumn{5}{c|}{\textbf{MNIST}} & \multicolumn{4}{c|}{\textbf{FashionMNIST}} & \textbf{Cifar-10}\\
      & (0,1) & (0,3)  & (2,4) & (5,6) & (2,8) & (0,1) & (2,8) & (3,9) & (7,9) & (0,1)  \\
      \hline
      Angle & 96.0 & 89.0 & 85.0 & 86.0 & 81.0 & 88.5 & 86.0 & 94.0 & 82.0 & 70.0  \\
      Amplitude & 95.5 & 88.5 & 84.0  & 85.5 & 79.5 & 88.0 & 84.5 & 87.0 & 78.0 & 68.5\\
      PCA & 99.0 & 88.0 & 84.5  & 85.0 & 86.0 & 88.5 & 86.0 & 93.0 & 79.0 & 68.0\\
      SQE & 88.0 & 86.0 & 82.0  & 83.5 & 78.5 & 86.0 & 83.0 & 81.0 & 77.0 & 66.0\\
      ATP & 99.0 & 91.0 & 86.0  & 87.0 & 83.0 & 91.5 & 86.0 & 94.0 & 83.0 & 74.2\\
      Ours & \textbf{99.8} & \textbf{99.6} & \textbf{98.8}  & \textbf{98.4} & \textbf{98.7} & \textbf{98.2} & \textbf{97.4} & \textbf{99.8} & \textbf{95.0} & \textbf{84.4}\\
      \hline
    \end{tabular}
  \end{center} 
\end{table*}
Across all pairwise binary classification tasks, FPQE consistently outperforms existing encoding strategies, with particularly large gains on more challenging dataset pairs and on the Cifar-10 benchmark.     On MNIST and FashionMNIST, most baselines achieve reasonable performance on simpler digit or texture distinctions, yet their accuracy begins to decline noticeably on pairs requiring fine-grained structural discrimination.     In contrast, FPQE maintains uniformly high performance across all pair combinations, suggesting that the fidelity-preserving representations retain the subtle variations necessary for reliable quantum decision boundaries.

The advantage of FPQE becomes even more pronounced on Cifar-10, where the visual complexity and texture diversity place much higher demands on the encoding stage.     While traditional methods exhibit a substantial drop in accuracy,  FPQE continues to achieve strong and stable performance.     This indicates that FPQE's ability to preserve spatial organization and local semantics directly translates into enhanced separability of the corresponding quantum states, enabling the QNN to distinguish between visually complex categories that challenge amplitude, angle, PCA, and SQE methods.

Overall, the results demonstrate that fidelity preservation is crucial to effective quantum learning, and that FPQE provides a robust, scalable strategy for bridging high-dimensional visual data with the constraints of NISQ era quantum models.     FPQE's consistent margins across multiple datasets and label pairs highlight its generality and its ability to provide structurally meaningful quantum inputs, particularly where baseline encodings struggle most.

\subsection{performance Results}
\begin{figure*}
  \centering
  \includegraphics[width=0.95\textwidth]{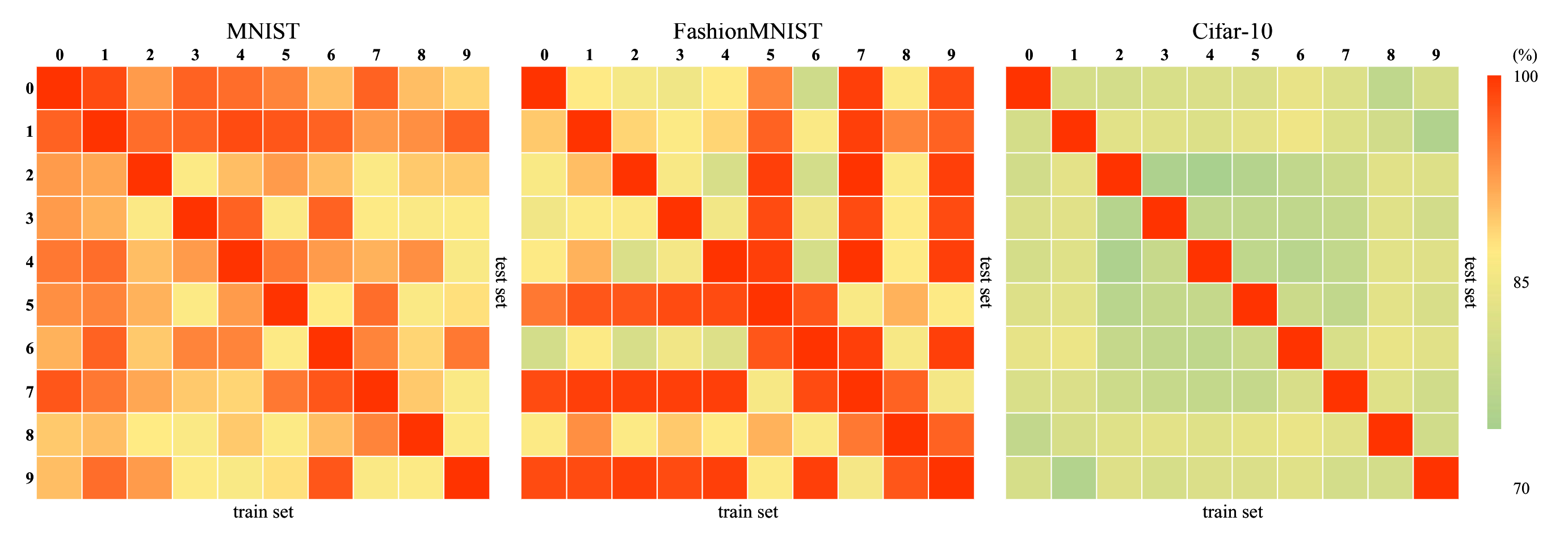}
  \caption{Visual results.}
  \label{visual_results}
\end{figure*}

In this subsection, we present a detailed analysis of the binary classification performance of FPQE across all labels in the MNIST, FashionMNIST, and Cifar-10 datasets.
The results, presented in Fig. \ref{visual_results}.

The full pairwise classification results reveal several important characteristics of FPQE. On MNIST and FashionMNIST, FPQE achieves consistently strong performance across nearly all label pairs. This stability indicates that FPQE preserves the essential structural cues, such as stroke continuity, local shapes, and garment contours, required for distinguishing between simple or moderately complex visual categories. These datasets contain relatively clean and well separated classes, and FPQE's fidelity-preserving representation allows the QNN to exploit these distinctions with minimal degradation, even under limited qubit budgets.

However, for pairs where different categories share similar textures or local appearance patterns, the classification performance drops compared to the datasets easier pairs.   These challenging cases reflect limitations not only of FPQE, but also of the quantum classifier's capacity to resolve subtle semantic differences when operating on compressed representations.

\subsection{Fidelity}
\textbf{Fidelity analyses}
  High fidelity data encoding is essential for enabling QNNs to capture the intrinsic correlations of classical data once mapped into quantum Hilbert space. Information loss during dimensionality reduction or quantum state preparation typically leads to degraded learning efficiency and reduced representational power. To assess the fidelity-preserving capability of our proposed framework, we conducted both quantitative and qualitative analyses comparing FPQE with conventional dimensionality reduction methods.


Our evaluation employs the decoded output $x'$ for computing indicator like MSE, PSNR and SSIM, primarily because the latent representations $z$ produced by different encoding methods are inherently incomparable, making a standardized assessment intractable.

Table \ref{fidelity_results} compares the reconstruction fidelity of different encoding schemes using MSE, PSNR, and SSIM across three datasets. While MSE and PSNR reflect pixel level differences, SSIM captures structural similarity, which is far more critical for quantum learning because structural degradation leads to highly overlapping quantum states after amplitude encoding.
\begin{table*}[htbp]
  \begin{center}
    \caption{The fidelity of different encoding methods.}
    \label{fidelity_results}
    \begin{tabular}{cc|ccc|ccc|ccc}
      \hline
      & Qubits & \multicolumn{3}{c|}{\textbf{MNIST}} & \multicolumn{3}{c|}{\textbf{FashionMNIST}} & \multicolumn{3}{c}{\textbf{Cifar-10}}\\
      \hline
      & & MSE & PSNR  & SSIM & MSE & PSNR  & SSIM & MSE & PSNR  & SSIM  \\
      \multirow{2}{*}{Angle} & 6 & 0.107 & 9.66 & 0.26 & 0.098 & 10.07 & 0.22 & 0.073 & 11.32 & 0.19  \\
      & 9 & 0.093 & 10.29 & 0.31 & 0.078 & 11.03 & 0.26 & 0.067 & 11.73 & 0.19  \\
      \multirow{2}{*}{Amplitude} & 6 & 0.034 & 14.62 & 0.61 & 0.029 & 15.24 & 0.54 & 0.021 & 16.6 & 0.37  \\
       & 8 & 0.010 & 19.65 & 0.88 & \textbf{0.011} & \textbf{19.24} & \textbf{0.79} & 0.008 & 20.46 & 0.63  \\
      \multirow{2}{*}{PCA} & 6 & 0.022 & 16.53 & 0.31 & 0.022 & 16.53 & 0.31 & 0.026 & 15.76 & 0.27  \\
      & 9 & 0.022 & 16.53 & 0.31 & 0.022 & 16.53 & 0.31 & 0.022 & 16.53 & 0.31  \\
      \multirow{2}{*}{SQE} & 6 & 0.095 & 10.21 & 0.25 & 0.092 & 10.36 & 0.21 & 0.068 & 11.65 & 0.19  \\
      & 9 & 0.093 & 10.29 & 0.31 & 0.078 & 11.03 & 0.26 & 0.067 & 11.72 & 0.19  \\
      ATP &  & - & - & - & - & - & - & - & - & -  \\
      Ours & 6 & \textbf{0.004} & \textbf{23.23} & \textbf{0.96} & 0.020 & 16.92 & 0.78 & \textbf{0.002} & \textbf{25.27} & \textbf{0.85}  \\
    \end{tabular}
  \end{center}
\end{table*}

Across MNIST and FashionMNIST, FPQE achieves substantial improvements over all baselines, but the difference becomes most evident on Cifar-10. Interestingly, although the MSE and PSNR of baseline encoders do not degrade significantly on Cifar-10, their SSIM values drop sharply (e.g., Angle: 0.19, Amplitude: 0.63, PCA: 0.27). This reveals a key observation: traditional encoders can reconstruct images with similar pixel intensities, but they fail to preserve local structures as data complexity increases. In other words, the reconstructions “look similar” in terms of raw values but lose important edges, textures, and semantic regions—information crucial for forming separable quantum states.

FPQE maintains both low MSE and high structural fidelity across all datasets with only 6 qubits. FPQE achieves 0.002 MSE, 25.27 dB PSNR, and 0.85 SSIM on Cifar-10, outperforming amplitude encoding by a large margin. The ability to retain high SSIM under tight qubit constraints demonstrates that FPQE preserves spatial relationships and fine grained textures rather than merely replicating average pixel values.

\begin{figure*}
  \includegraphics[width=0.9\textwidth]{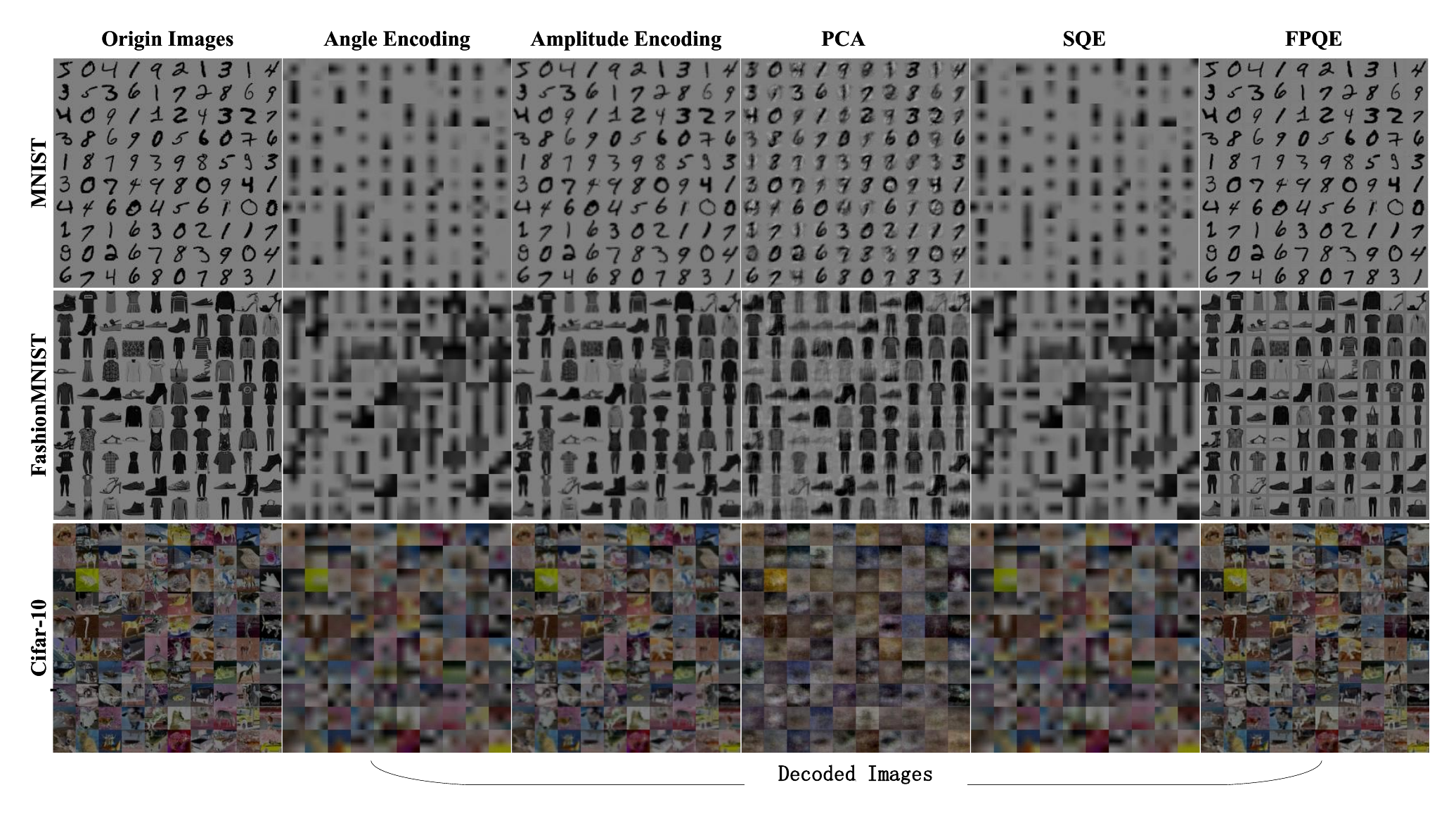}
  \caption{Visualize of fidelity}
  \label{visual_fidelity}
\end{figure*}

\textbf{Visual Fidelity: }Fig. \ref{visual_fidelity} compares reconstructed samples for MNIST and FashionMNIST. FPQE preserves contour sharpness and intra class texture details, while PCA and pooling exhibit feature loss due to linear or local averaging operations. The visual results align with fidelity metrics, validating FPQE's near lossless property.
\section{Discussion and Conclusion}
\label{sec:con}
\subsection{Discussion and Insights}
Our study highlights the critical role of fidelity preservation in classical to quantum data encoding. While prior work has focused primarily on reducing input dimensionality to fit NISQ hardware constraints, our findings demonstrate that structure preserving compression is far more important than dimensionality alone for downstream quantum learning performance. Experiments show that even when baselines achieve comparable MSE or PSNR, their severe SSIM degradation especially on complex datasets like CIFAR-10, leads to highly overlapping quantum states after amplitude encoding. This collapse in structural fidelity directly weakens the discriminative capacity of quantum neural networks.

FPQE addresses this issue by learning a compressed representation that maintains global layout, edge information, and textural patterns. As a result, the encoded vectors produce quantum states that remain more distinguishable in Hilbert space, enabling QNNs to leverage their expressive advantage. The growing performance gap between FPQE and classical encoders as data complexity increases further suggests that quantum models benefit disproportionately from structurally rich inputs, more so than from pixel accurate but structurally degraded features.

Another important observation is that FPQE enables effective quantum learning with fewer qubits. By mapping images to multi-channel low-dimensional tensors that still encode spatial structure, FPQE bypasses the need for extremely high qubit counts typically required for image level amplitude encoding. This hardware efficiency makes FPQE a practical choice for near term quantum devices and a foundation for scaling quantum representation learning in future high capacity quantum architectures.

\subsection{Limitations and Future Work}

Although FPQE achieves strong fidelity preservation and improved quantum classification performance, several limitations remain. First, our experiments focus solely on binary image classification, and FPQE has not yet been validated on more complex tasks such as multiclass recognition or structured prediction, where the benefits of structural fidelity may manifest differently. Second, the framework currently relies on convolutional priors tailored to visual data, limiting its applicability to domains such as natural language or graph structured information, which would require alternative encoders capable of capturing non-spatial structure. Third, FPQE is evaluated only in discriminative settings and has not been explored in quantum generative modeling, despite its potential as a stable latent representation for quantum VAEs or diffusion models. Finally, all evaluations are conducted on simulators, although FPQE uses relatively few qubits and should therefore be resilient to typical NISQ noise, real hardware may still introduce decoherence and amplitude loading errors that affect fidelity preservation. Future studies are needed to quantify these effects and adapt FPQE for practical deployment.\\
\textbf{Future work will explore three main directions:}
\begin{enumerate}
    \item \textbf{Expanding Beyond Classification.} We plan to extend FPQE to multiclass recognition, segmentation, and visual reasoning, examining whether fidelity preservation continues to benefit more complex decision boundaries.
    \item \textbf{Toward Quantum Generative Models.} FPQE's structurally stable latent space provides a natural foundation for quantum generative pipelines. Future exploration includes quantum VAEs, quantum GANs, and quantum diffusion models.
    \item \textbf{Distributed Quantum Execution.} As modular quantum architectures evolve, FPQE can be mapped onto distributed quantum chips, enabling larger quantum encodings and deeper QNNs. Such architectures could support coherent, measurement free execution, preserving quantum state continuity across the entire pipeline.
\end{enumerate}

\subsection{Conclusion}
In this work, we introduced FPQE, a fidelity-preserving encoding framework designed to bridge the gap between high-dimensional visual data and the limited representational capacity of NISQ quantum hardware. Unlike conventional classical to quantum encodings that prioritize dimensionality reduction at the cost of structural distortion, FPQE explicitly learns to preserve both global semantics and local textures through a convolutional encoder-decoder. By delivering compact yet structurally faithful representations, FPQE generates quantum states with higher separability, enabling downstream QNNs to exploit their expressiveness more effectively.

Extensive experiments across MNIST, FashionMNIST, and CIFAR-10 demonstrate that FPQE consistently achieves superior reconstruction fidelity and improved quantum classification performance. Notably, FPQE's advantage becomes more pronounced as data complexity increases, showing that structural fidelity is crucial for generalizable quantum learning. Our analysis further reveals that traditional encodings maintain pixel level similarity but experience sharp declines in structural metrics such as SSIM, explaining their degraded performance on challenging datasets.

FPQE provides a scalable, hardware feasible pathway for preparing quantum ready visual representations and highlights the importance of fidelity driven design in future quantum machine learning pipelines. We believe this direction opens new opportunities for integrating classical representation learning with quantum information processing and sets a foundation for robust quantum vision models as quantum hardware continues to advance.

{
    \small

}


\end{document}